\documentclass[aps,prl,showpacs,notitlepage,floatfix,twocolumn]{revtex4-1}
\usepackage{bbm}
\usepackage{mathrsfs}
\usepackage{epsfig}
\usepackage{graphicx}
\usepackage{amsfonts}
\usepackage[figuresright]{rotating}
\usepackage{amssymb}
\usepackage{amsmath}
\usepackage{dcolumn}
\usepackage{bm}
\usepackage{braket}
\usepackage{comment}
\usepackage{mathtools}
\usepackage{xcolor}
\usepackage{multirow}
\usepackage{setspace}

\def\avg#1{\langle#1\rangle}

\def\be{\begin{equation}} \def\ee{\end{equation}}
\def\bea{\begin{eqnarray}} \def\eea{\end{eqnarray}}

\def\nn{\nonumber}

\usepackage{color}

\usepackage[normalem]{ulem}

\begin{document}
\title{Two-band model for magnetism and superconductivity in
nickelates}
\author{Lun-Hui Hu}
\affiliation{Department of Physics, University of California,
San Diego, California 92093, USA}
\author{Congjun Wu}
\affiliation{Department of Physics, University of California,
San Diego, California 92093, USA}
\begin{abstract}
The recently discovered superconductivity in Nd$_{1-x}$Sr$_x$NiO$_2$
provides a new opportunity for studying strongly correlated
unconventional superconductivity.
The single-hole Ni$^+$ ($3d^9$) configuration in the parent compound
NdNiO$_2$ is similar to that of Cu$^{2+}$ in cuprates.
We suggest that after doping, the intra-orbital spin-singlet and
inter-orbital spin-triplet double-hole
(doublon) configurations of Ni$^{2+}$ are competing, and we
construct a two-band Hubbard model by including both
the $3d_{x^2-y^2}$ and $3d_{xy}$-orbitals.
The effective spin-orbital super-exchange model in the undoped case
is a variant of the $SU(4)$ Kugel-Khomskii model augmented by
symmetry breaking terms.
Upon doping, the effective exchange interactions between spin-$\frac{1}{2}$ single-holes, spin-1 (triplet) doublons, and singlet doublons are derived.
Possible superconducting pairing symmetries are classified in accordance
to the $D_{4h}$ crystalline symmetry, and their connections to the
superexchange interactions are analyzed.
\end{abstract}
\maketitle

The recent exciting discovery of the nickelate superconductivity
\cite{li2019}
has aroused a great deal of attention in the condensed matter community
\cite{hirsch2019,botana2019,sakakibara2019,hepting2019,nomura2019,gao2019,ryee2019,zhang2019b,wu2019,jiang2019,zhang2019b,zhang2019a}.
The infinite-layer nickelate, Nd$_{0.8}$Sr$_{0.2}$NiO$_2$, is synthesized
on the SrTiO$_3$ substrate, which
exhibits relatively high transition temperatures ($T_c$) around $9-15$ K.
The unusual electronic configuration of Ni$^{+}$ ($3d^9$) is similar to that of
Cu$^{2+}$ in high $T_c$ cuprate superconductors.
During the past three decades, the high $T_c$ superconductivity remains
one of the most outstanding problems in condensed matter physics \cite{bednorz1986,bednorz1988,anderson1987,anderson2004,lee2006}.
The parent cuprate compounds are charge-transfer insulators based on
the Zaanen-Sawatzky-Allen scheme \cite{zaanen1985} exhibiting the antiferromagnetic (AFM) long-range order.
Upon chemical doping, additional holes go to the oxygen
$2p$-orbitals, and are combined with the $3d_{x^2-y^2}$ spins of Cu$^{2+}$ cations to form the Zhang-Rice singlets
\cite{zhang1988}.
The $d$-wave superconductivity arises as doping suppresses the AFM
long range order \cite{shen1993,wollman1993,tsuei1994}.
It has been a long-lasting question if the fascinating physics in
high $T_c$ cuprates also exists in other strongly correlated oxides.
Due to its similarity to cuprates, a great deal of efforts
both experimental and theoretical have been made to investigate
the nickelate-based superconductivity \cite{anisimov1999,hayward1999,lee2004,chaloupka2008,
hansmann2009,boris2011,benckiser2011,disa2015,middey2016,botana2017,zhang2017}.

Although the nickelates exhibit a similar configuration to cuprates,
their behaviors are very different.
The most noticeable distinction is that no evidence of magnetic ordering
is experimentally observed \cite{hayward1999,hayward2003}.
One possible reason is the large charge transfer energy, i.e., the energy
difference between a single hole lying on the nickel and oxygen sites,
$\Delta_{pd} \approx 9$ eV in NdNiO$_2$, which is much larger
than $\Delta_{pd} \approx3$ eV in cuprates.
The superexchange energy scale is estimated as $J \sim 1/\Delta^2_{pd}$ \cite{jiang2019},
which is about one order smaller than that in cuprates \cite{zaanen1987},
and thus the AFM ordering is weakened.
Another possibility is the self-doping effect \cite{sakakibara2019,zhang2019b} to the Mott
insulating state, where the forming of Kondo singlets suppresses the
AFM ordering \cite{zhang2019a}.
The lattice constant along the $z$-axis in NbNiO$_2$ is only about
$3.4$ \AA, much smaller than that in cuprates.
This leads to the dispersion along the $z$-axis from the Nd $5d_{z^2}$-orbital \cite{ikeda2016,li2019}.
The Nd-originated electron pockets are found by the LDA+U calculations in  previous works \cite{anisimov1999,lee2004,chaloupka2008} and also in
recent works \cite{botana2019,sakakibara2019,hepting2019,nomura2019,gao2019,
ryee2019,zhang2019b,wu2019}.
Based on Ref. \cite{sakakibara2019}, the electron pocket volume
from Nd-electrons is estimated smaller than $4\%$ of the Brillouin
zone, and then the self-doping effect should be weak.
Due to the large charge transfer energy in nickelates, the extra holes
from self-doping also appear on Ni-sites, hence,
we construct a Ni-only model as a starting point.

Due to the large charge transfer energy in NdNiO$_2$,
extra holes by doping are commonly believed to lie on the
Ni-sites forming the double-hole configuration of Ni$^{2+} (3d^8$).
This is in sharp contrast to cuprates in which the doped holes
lie on oxygens.
So far most work on the nickelate superconductivity view
the Ni site as orbital-inactive -- only the $d_{x^2-y^2}$ orbital is
occupied for both the single-hole configuration of Ni$^+$ ($3d^9$)
and the doublon configuration of Ni$^{2+}$ ($3d^8$)
(Here and after, we use ``doublon'' for the double-hole configurations
of Ni$^{2+}$ following the convention in literatures.)
In other words, Ni$^{2+}$ is often assumed to be a spin singlet.

In this article, we examine the orbital property of
the Ni-site and its role in quantum magnetism and
superconductivity in nickelates.
Due to Hund's coupling and the relatively small inter-orbital repulsion,
the triplet doublon of the Ni$^{2+}$ cation is a competing configuration,
in which both the $d_{x^2-y^2}$ and $d_{xy}$-orbitals are occupied.
Based on the two-band Hubbard model, we study the effective
super-exchange processes among spin-$\frac{1}{2}$ single-holes,
triplet doublons, and singlet doublons.
Superconductivity arises due to the Cooper pairing between
doublons, which actually only carry unit charge.
Both singlet and triplet Cooper pairings can take place based on
different doublon configurations.
We also classify the possible pairing symmetries based on the
crystalline symmetry of nickelates.

\begin{figure}[!htbp]
\centering
\includegraphics[width=0.9\linewidth]{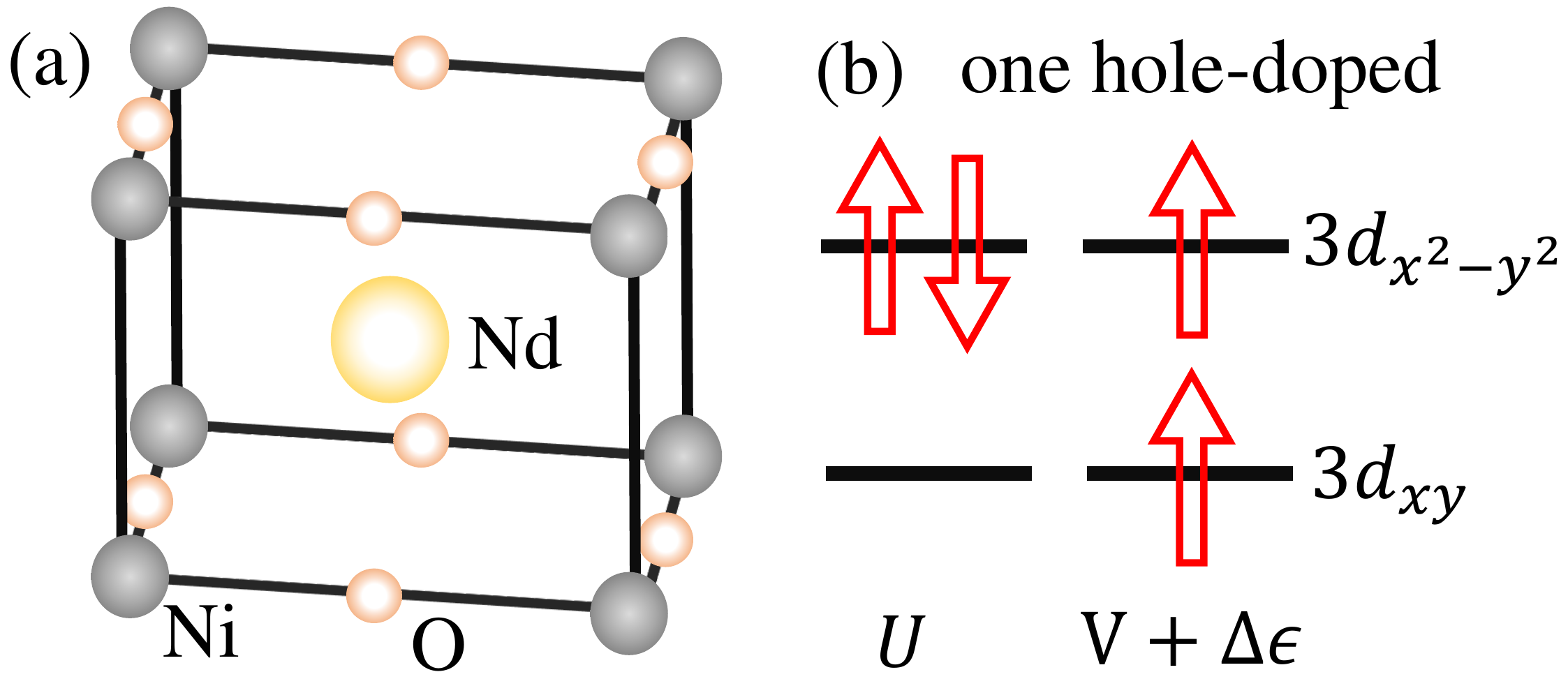}
\caption{($a$) The crystalline structure of NdNiO$_2$.
The space group symmetry is $P4/mmm$.
($b$) The competing singlet and triplet doublon (two-hole)
configurations of Ni$^{2+}$.
$U$ and $V$ are the intra- and inter-orbital interactions
defined in Eq. \ref{eq:hubbard}.
The hollow arrows represent two spin configurations of holes.
}
\label{fig:config}
\end{figure}

The parent compound NdNiO$_2$ possesses the space group crystalline symmetry
of $P4/mmm$.
The Ni-site is surrounded by four O$^{2-}$ anions forming a planar square
structure as shown in Fig.~\ref{fig:config}($a$), and the Ni$^+$ cation is
in the $3d^9$ configuration, i.e., a single hole on the Ni site.
The crystal field splitting of the five-fold $3d$-orbitals of Ni can be
intuitively analyzed as follows based on the tetragonal symmetry:
Due to the absence of the apical oxygen anions, the $d_{r^2-3z^2}$-orbital
extending along the $z$-direction has the lowest energy.
It is followed by the doubly degenerate $d_{xz}$ and
$d_{yz}$-orbitals, which also extend along the $z$-direction.
In contrast, the energies of the in-plane orbitals are pushed higher
by the negatively charged oxygen anions located in the middle of the
Ni-Ni bonds:
The $d_{x^2-y^2}$-orbital has the highest energy since it points
to the oxygen anions, followed by the $d_{xy}$-orbital
which extends along the diagonal direction of the NiO$_2$ plane.
The on-site energy difference between two highest $d_{x^2-y^2}$ and
$d_{xy}$-orbitals is estimated as $\Delta \varepsilon\sim 1.38 eV$
in Ref.  [\onlinecite{wu2019}].
Hence, without doping, the single hole lies in the $d_{x^2-y^2}$-orbital.

Now consider the doublon configuration when extra holes are doped.
At low and intermediate levels of doping, only doublons need to
be considered, and we neglect the small possibilities of three- and
four-hole configurations.
By keeping two highest electron orbitals, {\it i.e.}, two lowest energy
orbitals of holes, two competing doublon configurations, i.e.,
the singlet doublon only occupying the $d_{x^2-y^2}$-orbital,
and the triplet doublon occupying both $d_{x^2-y^2}$ and $d_{xy}$-orbitals,
are shown in Fig.~\ref{fig:config} ($b$).
The standard onsite two-orbital Hubbard interactions plus the
onsite single-hole energy term read
\begin{eqnarray}
\mathcal{H}_{int} (i) &=& \Delta \varepsilon n_2(i)+ U\sum_{a=1,2}
n_{a\uparrow} (i) n_{a\downarrow} (i)
+V n_1(i) n_2 (i) \nn\\
&-&J \left( \vec{S}_{1}(i) \cdot\vec{S}_{2} (i)-
\frac{1}{4} n_{1}(i) n_{2}(i) \right),
\label{eq:hubbard}
\end{eqnarray}
where $i$ is the site index;
$a=1$ and 2 represent the $d_{x^2-y^2}$ and $d_{xy}$-orbitals, respectively;
$n_a$ and $\vec S_a$ are the hole-number and hole spin operators in orbital $a$,
respectively;
$U$ is the intra-orbital interaction strength,
$J$ is Hund's coupling, and $V$ is the
inter-orbital interaction.
Due to the relatively large splitting $\Delta\varepsilon$ between
these two orbitals, the pairing hopping interaction is neglected.
The energies of the spin singlet and triplet doublons are
$U$ and $V+\Delta\varepsilon$, respectively.
Using the estimations in Ref. [\onlinecite{sakakibara2019}],
$U=3.8eV$, $V=1.9eV$, $J=0.7eV$ ($U^\prime=V+J=2.6eV$ in their convention.),
the triplet energy, $V+\Delta\epsilon$, is smaller than that of the singlet.
Nevertheless, accurate estimations on these parameters are very difficult
at this early stage of research.
Considering the uncertainty, it is reasonable to assume that their energies are close.
The near degeneracy of the above competing configurations motivates
us to employ the two-orbital model to describe magnetism and
superconductivity in nickelates.

We define the two-band Hubbard model as
$\mathcal{H}=\mathcal{H}_t+\sum_i \mathcal{H}_{int}(i)$, with the
hopping Hamiltonian given below as
\begin{align}
\mathcal{H}_{t} = -\sum_{\langle ij\rangle}\sum_{a=1,2}\sum_{\sigma=\uparrow,\downarrow}
t_a \hat{c}_{a,\sigma}^\dagger(i) \hat{c}_{a,\sigma}(j) + \text{h.c.},
\end{align}
where $t_{1,2}$ are the nearest-neighbor (NN) intra-orbital hoppings
with $1(2)$ representing the $d_{x^2-y^2}$ and $d_{xy}$-orbitals, respectively.
$t_1$ is expressed as $t_1=t_{1,dd} + (t_{1,pd})^2/\Delta_{pd}$,
where $t_{1,dd}$ is the direct overlap between $d_{x^2-y^2}$ orbitals on
neighboring Ni-sites, and the second term describes the assisted hopping
via the oxygen $2p$-orbital.
Due to the large charge transfer energy $\Delta_{pd}$ in nickelates,
these two contributions are comparable.
Similar analysis can be performed to $t_2$.
The NN inter-orbital hoppings are forbidden due to the different
symmetries of the $d_{x^2-y^2}$ and $d_{xy}$-orbitals.
For simplicity, we neglect the difference between these two hopping
integrals by setting $t_1=t_2=t_0$.

\begin{table}[!htbp]
\renewcommand\arraystretch{1.3}
		\setlength{\tabcolsep}{3.5mm}
	\begin{ruledtabular}
		\centering
		\begin{tabular}{c|c|c}
			$D_{4h}$             & $\psi(\bf{k})$     & $M$  \\ \cline{1-3}
	\multirow{2}{*}{$A_{1g}$}    & $ \cos k_x+ \cos k_y, ~\cos k_z$      & $\tau_0,\tau_z$            \\
	                             & $\sin k_x\sin k_y(\cos k_x-\cos k_y)$      & $\tau_x$                   \\ \hline
\multirow{2}{*}{$A_{2g}$}    & $\sin k_x\sin k_y(\cos k_x- \cos k_y)$      & $\tau_0,\tau_z$            \\
	                             & $\cos k_x+\cos k_y,~\cos k_z$      & $\tau_x$                   \\ \hline
    \multirow{2}{*}{$B_{1g}$}    & $\cos k_x- \cos k_y$              & $\tau_0,\tau_z$            \\
                                 & $\sin k_x \sin k_y$                   & $\tau_1$                   \\ \hline
	\multirow{2}{*}{$B_{2g}$}    & $\sin k_x \sin k_y$                   & $\tau_0,\tau_z$            \\
	                             & $\cos k_x- \cos k_y$              & $\tau_x$                   \\ \hline
			$E_{g}$              & $(\sin k_x\sin k_z, ~\sin k_y\sin k_z)$          & $\tau_0,\tau_x,\tau_z$     \\
	\end{tabular}
\end{ruledtabular}
\caption{The form factors of $\psi(\bf{k})$ for the even-parity,
spin-singlet and orbital-symmetric gap functions in Eq. \ref{eq:gap_1}.}
\label{tab:singlet_even}
\end{table}

Superconducting gap function symmetries are a central problem
of unconventional superconductivity.
At the current stage, this problem remains difficult for nickelate
superconductors.
Below we classify gap function symmetries enriched by the multi-orbital
structure to provide guidance for later research.
These symmetries according to the $D_{4h}$ point group representations \cite{sigrist1991}
are $A_{1g(u)}$, $A_{2g(u)}$, $B_{1,g(u)}$, $B_{2,g(u)}$
and $E_{g(u)}$, where the subscript $1 (2)$ represents the even (odd) parity
of the reflection with respect to the $xy$ or $yz$-planes;
$g (u)$ denotes the even (odd) parity with respect to inversion;
$A$, $B$, and $E$ indicate the discrete orbital angular momenta
of $0$, $2$, and $\pm 1$, respectively.
We first consider the  gap functions in the spin-singlet channel.
They are represented in the two-orbital formalism as
\begin{eqnarray}
\hat{\Delta} (\mathbf{k}) =  \psi({\mathbf k}) i\sigma_y M,
\label{eq:gap_1}
\end{eqnarray}
where $i\sigma_y$ is the charge conjugation matrix, and $M$ is the orbital
pairing matrix.
We use $\tau_{x,y,z}$ for the Pauli matrices in the orbital channel, where $\tau_z=\pm\frac{1}{2}$ refers to the $d_{x^2-y^2}$
and $d_{xy}$-orbitals, respectively, and $\tau_0$ is the identity matrix.
The Fermi statistics imposes the constraint $\hat{\Delta}(\mathbf{k}) = - \hat{\Delta}^T (-\mathbf{k})$.
The gap functions in the spin-singlet even-parity channels are listed
in Tab. \ref{tab:singlet_even}
 in which $\tau_{0,z}$ refer to
the intra-orbital pairing, and $\tau_x$ refers to the inter-orbital
symmetric pairing.
Considering the possible ferromagnetic fluctuations in nickelate
superconductors, we also consider the triplet pairings,
\begin{eqnarray}
\hat{\Delta} (\mathbf{k}) =  \mathbf{d}(\mathbf{k})  \cdot\bm{\sigma}
i\sigma_y M,
\label{eq:gap_2}
\end{eqnarray}
where $\mathbf{d}(\mathbf{k})$ is the so-called $d$-vector for triplet
superconductors.
The odd-parity triplet pairing gap functions are listed in
Tab. \ref{tab:triplet_odd}.
Due to the orbital-dependence, gap functions can also be
odd-parity spin-singlet and even-parity spin-triplet,
nevertheless, they are unlikely to be relevant to the nickelate
superconductivity based on the analysis later
in this article.
The hoppings of the $d_{x^2-y^2}$ and $d_{xy}$-orbitals along the $z$-direction
are small, nevertheless, for completeness, we still keep the
$k_z$-dependent gap functions.

\begin{table}[!htbp]
	\renewcommand\arraystretch{1.3}
	\setlength{\tabcolsep}{3.5mm}
	\begin{ruledtabular}
		\centering
		\begin{tabular}{c|c|c}
			$D_{4h}$             & $\bf{d}(\bf{k})$     & $M$\\ \cline{1-3}
			\multirow{2}{*}{$A_{1u}$}    & $\sin k_z\hat{z},~\sin k_x\hat{y}+\sin k_y\hat{x}$      & $\tau_0,\tau_z$        \\
			                             & $\sin k_y\hat{y}-\sin k_x\hat{x}$                 & $\tau_x$               \\ \hline
			\multirow{2}{*}{$A_{2u}$}    & $\sin k_y\hat{y}-\sin k_x\hat{x}$                 & $\tau_0,\tau_z$        \\
			                             & $\sin k_z\hat{z}, ~\sin k_x\hat{y}+\sin k_y\hat{x}$      & $\tau_x$               \\ \hline
			\multirow{2}{*}{$B_{1u}$}    & $\sin k_x\hat{y}-\sin k_y\hat{x}$                 & $\tau_0,\tau_z$        \\
			                             & $\sin k_y\hat{y}+\sin k_x\hat{x}$                 & $\tau_x$               \\ \hline
			\multirow{2}{*}{$B_{2u}$}    & $\sin k_y\hat{y}+\sin k_x\hat{x}$                 & $\tau_0,\tau_z$        \\
			                             & $\sin k_x\hat{y}-\sin k_y\hat{x}$                 & $\tau_x$               \\ \hline
			$E_{u}$              & $(\sin k_x,\sin k_y)\hat{z}, ~~\sin k_z(\hat{x},\hat{y})$ & $\tau_0,\tau_x,\tau_z$     \\
		\end{tabular}
\end{ruledtabular}
\caption{The $d$-vectors for the spin triplet, odd-parity, and orbital
symmetric pairing gap functions in Eq. \ref{eq:gap_2}.}
\label{tab:triplet_odd}
\end{table}

Now we consider the strong coupling aspect of the nickelate
physics.
The effective model in the strong interaction limit is constructed below
via the 2nd order perturbation theory.
The undoped configuration corresponds to the 1/4-filling of holes in two bands,
{\it i.e.}, a spin-$\frac{1}{2}$ hole on each site.
The effective super-exchange model of the bond $\avg{ij}$ can be
derived as
\begin{eqnarray}
H_{ex}(ij)&=&
-\Delta \varepsilon \Big( \tau_z(i)+\tau_z(j) \Big)
-J_{FM} P_t^s(ij) P_{s}^o(ij) \nn \\
&-&J_{AF} P_s^s(ij) (\tau_z(i)+\frac{1}{2}) (\tau_z(j)+\frac{1}{2}),
\label{eq:exchange}
\end{eqnarray}
where $P_t^s=\vec S(i) \cdot \vec S(j)+\frac{3}{4}n(i) n(j)$ and
$P_s^s=-\vec S(i) \cdot \vec S(j)+\frac{1}{4}n(i)n(j)$ are the
projection operators to the bond spin triplet and
singlet sectors, respectively, with $n(i)$ the hole number
on site $i$;
$P^o_s= -\vec \tau_i \cdot \vec\tau_j+\frac{1}{4}n(i)n(j)$
is the projection operator in the orbital singlet channel.
The orbital-flipping super-exchange process is ferromagnetic (FM)
represented by $J_{FM}\approx 4t_0^2/V$, while the spin-flipping
super-exchange process is represented by $J_{AF}\approx 4t_0^2/U$ is AFM .
Due to the onsite energy splitting, only the configuration
with both holes in the $d_{x^2-y^2}$-orbital is taken
into account in the AFM super-exchange.
Eq. \ref{eq:exchange} is a variant of the SU(4) Kugel-Khomskii
spin-orbital model augmented by symmetry breakings \cite{kugel1982,li1998}.
Consider an AFM ordered N\'eel configuration with all holes lying
in the $d_{x^2-y^2}$-orbital.
Let us flip a hole's spin and put it into the $d_{xy}$-orbital.
The exchange energy gain is roughly $\Delta E_{ex}=2zt_0^2(\frac{1}{V}
-\frac{1}{U})$ with $z=4$ the coordination number.
It is difficult to precisely estimate $\Delta E_{ex}$.
It should  be significantly smaller than $\Delta \varepsilon$,
nevertheless, conceivably, they are still at the same order.
Hence, the AFM ordering tendency would be significantly reduced,
which is in agreement with the absence of AFM long range order
in experiments.
Nevertheless, in the undoped case, hole remains in the
$d_{x^2-y^2}$-orbital as described by
\begin{eqnarray}
H_{ex}(ij)&=& \tilde{J}_{AF} \Big(\vec S(i) \cdot \vec S(j)
-\frac{1}{4}n_i n_j \Big),
\label{eq:exchange_reduced}
\end{eqnarray}
where $\tilde{J}_{AF}$ represents the reduced AFM exchange
by orbital fluctuations.

\begin{figure}[!htbp]
\centering
\includegraphics[width=\linewidth]{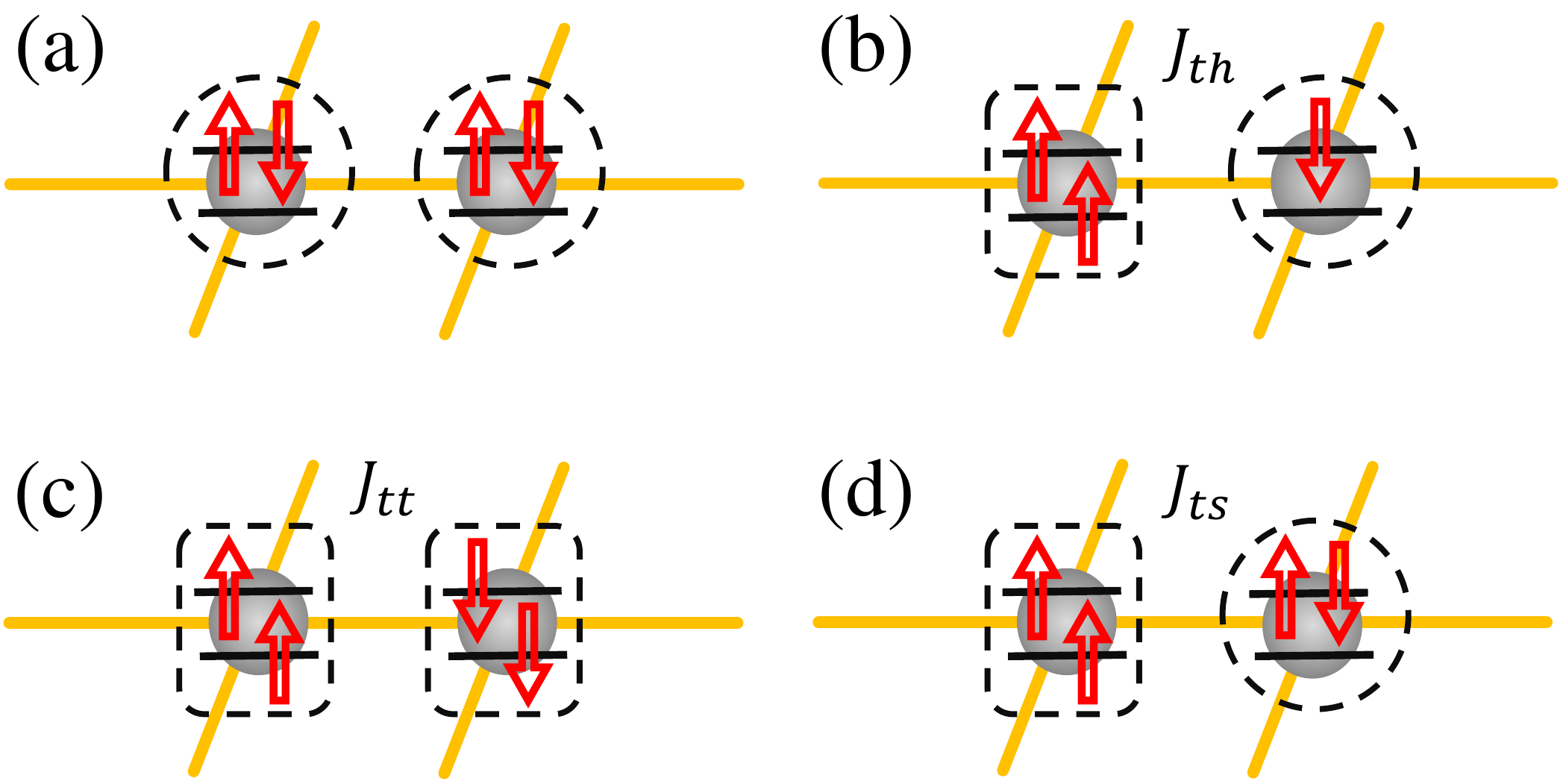}
\caption{Bond configurations with doublons and single-holes.
($a$) Two neighboring singlet doublons.
($b$) A triplet doublon and a spin-$\frac{1}{2}$ single hole.
($c$) Two neighboring triplet doublons.
($d$) A triplet doublon and a singlet doublon.
}
\label{fig:exchange}
\end{figure}

Next we construct the low energy super-exchange Hamiltonians after doping,
which include both doublons and single holes, via the 2nd order perturbation theory.
The possibility of a single hole in the $d_{xy}$-orbital is neglected
for the onsite energy splitting.
There is no exchange interaction between two neighboring singlet doublons
(Fig. \ref{fig:exchange} ($a$)), and no exchange interaction
between a singlet doublon and a single spin-$\frac{1}{2}$ hole, either.
In contrast, between a triplet doublon and a neighboring single
spin-$\frac{1}{2}$ hole shown in Fig. \ref{fig:exchange} ($b$),
the super-exchange interaction is ferrimagnetic,
\bea
H^{th}_{ex}(ij)=J_{th}\Big(\vec {T}_i\cdot \vec S_j -\frac{1}{4} n_i n_j\Big),
\label{eq:dh}
\eea
where ${\vec T}$ is the spin-1 operator of the triplet doublon,
and $J_{th}=\frac{3t^2}{2}(\frac{1}{U-V}+\frac{1}{U+V+J/2})$.
Furthermore, the superexchange interaction also exists
between two neighboring triplet doublons (Fig. \ref{fig:exchange} ($c$)),
which is described by the spin-1 AFM Heisenberg model as
\bea
H^{tt}_{ex}(ij)=J_{tt} \Big(\vec T_i \cdot \vec T_j-\frac{1}{4}n_i n_j\Big),
\label{eq:dd}
\eea
with $J_{tt}=\frac{2t^2}{U+J/2}$.
Finally, if we bring a triplet and a singlet doublons together
(Fig. \ref{fig:exchange} ($d$)),
their exchange interaction is described by
\bea
&&H^{ts}_{ex}(ij)=
-J_{ts} \Big(d_m^\dagger(i) d_0(i) d_0^\dagger(j) d_m(j)
+h.c.\Big) \nn \\
&+&J_{ts}
\Big(d_m^\dagger(i) d_m(i) d_0^\dagger(j) d_0(j)
+  d_0^\dagger(i) d_0(i) d_m^\dagger(j) d_m(j)\Big)
\nn
\eea
where $d^\dagger_{0,\pm1}$ are creation operators for doublons
and $J_{ts}=\frac{4t^2}{V+J/2}$.
In addition to superexchange interactions, a doublon and a single
hole can switch their positions.
For the singlet doublon and a single hole, it is simply a
straightforward hopping process.
The switching between a triplet doublon and a single hole
is described by
\bea
H^{th}_t(ij) &=& -t^\prime\sum_{m\sigma;m^\prime\sigma^\prime}
\Bigg\{
\left\langle jj_z\Big{|}1m\frac{1}{2}\sigma\right\rangle
\left\langle jj_z\Big{|}1m^\prime\frac{1}{2}\sigma^\prime\right\rangle
\nn \\
&\times& d^\dagger_m(i) c^\dagger_{1\sigma}(j)
 c_{1\sigma^\prime}(i) d_{m'}(j) +h.c.
 \Bigg\},
\eea
where $\langle .|.\rangle$ are the Clebsch–Gordan coefficients
between spin-1 and spin-1/2 sectors, and $t^\prime$ is
at the same order of $t_0$.

Now consider the glue for superconductivity based on the
above superexchange picture.
Compared to the undoped case, the doublon charge is one instead of two
compared to the background of single holes.
Pairing of two doublons leads to superconductivity.
An effective attraction between two singlet doublons at neighboring
sites is at the energy scale of $\Delta E_{ss}=-\frac{1}{2}\tilde{J}_{AF}$.
Bringing two neighboring triplet doublons together, the energy difference
compared to when they are apart is $\Delta E_{tt}=2 J_{th}-2J_{tt} -\frac{1}{2} \tilde{J}_{AF}$.
Finally, bringing a triplet and a singlet doublons together,
the energy difference is $\Delta E_{ts}= J_{th}-J_{ts}
-\frac{1}{2}\tilde{J}_{AF}$.
When each of the above quantities becomes negative, it means an effective attraction
in the corresponding channel.
Both cases of two singlet doublons and two triplet doublons can form spin
singlet Cooper pairing, and the pairing of a singlet doublon
and a triplet doublon gives rise to a triplet pairing
superconductivity.
Due to the relatively large value of $J_{th}$, the pairing strength
of triplet doublons is weak if not completely suppressed.

Next we connect the above doublon pairing picture to the
previous analysis on gap function symmetries.
Consider a Ni-Ni bond $\avg{ij}$.
In the absence of doping, we take the bond singlet state with one
hole in the $d_{x^2-y^2}$-orbital on each site as the background
state $|\Psi_0\rangle$.
Furthermore, the state of two singlet doublons (Fig.~\ref{fig:exchange}($a$))
is denoted as $|\Psi_{ss}\rangle$, and that of two triplet doublons
(Fig.~\ref{fig:exchange}($c$)) is denoted as $|\Psi_{dd}\rangle$.
These states can be connected to $|\Psi_0\rangle$ via the pairing operators
 $\chi_{ij}^{\dagger,ss (dd)} =\frac{1}{\sqrt 2} (\hat{c}_{a\uparrow}^\dagger(i) \hat{c}_{a\downarrow}^\dagger(j) - \hat{c}_{a\downarrow}^\dagger(i) \hat{c}_{a\uparrow}^\dagger(j))$ with $a=d_{x^2-y^2}$
for  $\chi_{ij}^{\dagger,ss}$ and $d_{xy}$
for $\chi_{ij}^{\dagger,dd}$, such that
$\langle \Psi_{ss (dd)}\vert \chi_{ij}^{\dagger,ss(dd)} \vert
\Psi_0\rangle\neq0$.
Then the orbital pairing matrices for $\chi_{ij}^{\dagger,ss (dd)}$
correspond to $\tau_0\mp \tau_z$, respectively.
According to Tab. \ref{tab:singlet_even}, the plausible pairing
symmetries are $A_{1g}$ ($s$-wave), and $B_{1g}$ ($d_{x^2-y^2}$-wave),
with the form factors  $\cos k_x \pm \cos k_y$, respectively.
As for the triplet pairing between singlet and triplet doublons, the bonding
state corresponds to the odd-parity combinations of the configuration
in Fig. \ref{fig:exchange} ($d$) and its parity partner.
The corresponding pairing operators are orbital symmetric and odd-parity:
$A_{1u}$, $A_{2u}$, $B_{1u}$, $B_{2u}$, $E_u$.
Their $d$-vector configurations exhibit the $p$-wave orbital symmetries of
$\sin k_x$ and $\sin k_y$ are shown in Tab. \ref{tab:triplet_odd}.

Energetically, it is more favorable if the gap function nodes are
away from the van Hove singularities of density of states.
Current band structure calculations show large density of states
around $(0,\pi)$ and $(\pi,0)$.
Hence, the $B_{1g}$ ($d_{x^2-y^2}$) singlet pairing symmetry is
probably the dominant one.
The singlet $A_{1g}$ and $p$-wave triplet pairings are competing
pairing symmetries but less favorable.
Since the singlet doublon pairing force is stronger, the
$d_{x^2-y^2}$-orbital pairing is expected to be more dominant
than that in the $d_{xy}$-orbital.

We note that a similar two-band model based on the nearly degenerate $e_g$-orbitals of $d_{x^2-y^2}$ and $d_{r^2-3z^2}$ was recently developed
for CuO$_2$ monolayers \cite{jiang2018}, as well as the high $T_c$ cuprate superconductor Ba$_2$CuO$_{3+x}$ \cite{le2019}. In nickelates the planar $B_{2g}$-orbital $d_{xy}$ replaces the $z$-directional $d_{r^2-3z^2}$ orbital in the CuO$_2$ plane whose symmetry is reduced to $A_{1g}$ under the tetragonal symmetry. Due to the different orbital symmetries, the inter-orbital pairing symmetries are different in these two classes of systems.

{\it Conclusion--.}
We suggest that the low-energy physics in Nd$_{1-x}$Sr$_x$NiO$_2$
is captured by the two-orbital model.
The second-order perturbation theory is employed to derive the effective Hamiltonians in both cases with and without dopings.
In the absence of doping, it is a variant of the $SU(4)$ Kugel-Khomskii
spin-orbital model subject to symmetry breaking terms.
It shows the competition between FM and AFM exchanges,
thus the magnetic tendency is significantly suppressed.
When additional holes are doped into nickelates,
two competing configurations of Ni$^{2+}$ appear: the intra-orbital
singlet doublon and the inter-orbital triplet doublon.
The superexchange interactions among two types of doublons and single-holes
are derived, and doublon pairings are studied
in the superexchange picture.
Possible pairing symmetries are analyzed based on the $D_{4h}$
point group.

{\it Acknowledgments--.}
C. W. thanks J. Hirsch for helpful discussions.
L. H. and C. W. are supported by AFOSR FA9550-14-1-0168.

{\it Note Added--.} Near the completion of the manuscript,
we became aware of the references arXiv:1909.12865 \cite{zhang2019c} and arXiv:1910.00473 \cite{werner2019},
in which the triplet doublons of the Ni$^{2+}$ cations are also proposed.

\bibliographystyle{apsrev4-1}
\bibliography{ref}

\end{document}